# Attributes of flares in Gamma Ray Bursts: sample I[*]

G. CHINCARINI[1,2], A. MORETTI[2], P. ROMANO[1,2], A.D. FALCONE[3], S. CAMPANA[2], S. COVINO[2], C. GUIDORZI[1,2], G. TAGLIAFERRI[2], D.N. BURROWS[3], M. CAPALBI[4], G. CUSUMANO[5], P. GIOMMI[4], , V. MANGANO[5], V. LA PAROLA[5], T. MINEO[5], M. PERRI[4], E. TROJA[5,6].

(1) Universita` degli Studi di Milano, Bicocca, Piazza delle Scienze 3, I-20126, Milano, Italy
(2) INAF – Osservatorio Astronomico di Brera, Via E. Bianchi 46, I-23807, Merate, LC, Italy
(3) Department of Astronomy and Astrophysics, Pennsylvania State University, University Park, PA 16802, USA
(4) ASI Science Data Center, ESRIN, Via G. Galilei, I-00044, Frascati, Italy
(5) INAF – Istituto di Fisica Spaziale e Fisica Cosmica sezione di Palermo, Via U. La Malfa 153, I-90146, Palermo, Italy.
(6) Department of Physics and Astronomy, University of Leicester, Leicester, UK

**Summary**

We discuss some of the preliminary results and findings derived from the analysis of a first sample of flares detected by the XRT on board Swift. The analysis shows that the morphology of flares is the one we expect from the collision of ultra-relativistic shells as it happens during the internal shock model proposed by Rees and Meszaros in 1994. Furthermore the $\frac{\Delta t}{t}$ ratio and the decay-time to rise-time ratio have mean values that are in good agreement with the values observed in the prompt emission pulses that are believed to originate from internal shocks. The conclusion is that the flare analysis favors the internal shock as due to shells that have been ejected by the central engine after the prompt emission. The central engine seems to remain active and capable of generating large amounts of energy also at later times. More data are needed to establish whether or not some of the flares could be due to shells that have been emitted with small Lorentz factor at the time of the prompt emission and generate later time flares due to the catch up of these shells.

1. **Introduction.**

The Swift Mission [1] changed drastically our approach to the study of Gamma Ray Bursts (GRBs) and their connection to Supernovae. From a field dominated by theory and a very few discoveries we passed, as it is characteristics of any astronomical field, to a field led by observations. Perhaps the most impressive discovery has been the location of short GRBs, the solution of a mystery that puzzled the astronomical community since their discovery. Basic discoveries on the way the phenomenon evolves have been due to the multi wavelength capabilities of the on board instrumentation and to the unique capability of the spacecraft in re-pointing the satellite in less than 100 seconds and derive an arcsec position after the BAT instrument has derived (in less than 15 seconds) the position to arc minutes accuracy. This has been reviewed in various meetings and we refer to Gehrels [2] (these proceedings) for an updated discussion of the most recent results.

---
[*] Paper presented at the GRB meeting in Venice: Swift and GRBs: Unveiling the relativistic Universe.

The large amount of data collected, on the other hand, pose questions related to the previous theoretical work and at the same time is allowing us to follow the events in a more empirical way as to single out those characteristics that best may lead us to the understanding of the physics of the explosion, to its evolution and to all those matters that are related to such phenomenon. Indeed the goal is to understand the characteristics of the progenitors also in relation to stellar evolution, the parent population of the progenitors and the role of the metallicity. Do GRBs track the cosmic star formation or rather the formation of metals. The Swift mission opened a new window into a cosmological world that we thought we knew but that we are re-discovering with new characteristics. We use this introduction to mention a few of the most intriguing problems that we are confident we will be able to solve with the data we are collecting. These are parts of a mosaic that we are building piece by piece.

Long and short bursts are two classes of bursts that seem to be due to a different progenitor and quite different mechanism of formation, collapsar or NS – NS or BH – NS merging, with a common output however: black hole and accretion disk. Such empirical classification is largely instrument dependent and under various aspects unsatisfactory. Recently this has been clearly outlined and a better classification may be based on the spectral lag [3]. On the other hand the recent observations show that the situation may be more complicated. The crisis of the classification exploded with the observations of GRB060614. Following the discovery of the absence of the expected Supernova [4], we looked into the characteristics of the prompt emission. We discovered that this burst had both the characteristics of a long GRB, $T_{90} = 102$ s with a long high energy emission after the first pulse, and of a short burst, null spectral lag [3]. The prompt emisson light curve observed for this burst furthermore is similar to the variability observed in the prompt emission of other GRBs, in particular it is very similar to what we observe in GRB050724.

GRB060614 to some extent focuses on the problem of the connection between GRBs and SNe and indirectly to the mechanism of the burst and of the related evolution. The picture we propose, in agreement with our previous work [4], is that the SN associated to a GRB could be visible or not depending on the mass of the newly formed black hole. Indeed it is well known that the collapse of a massive nucleus would lead to a massive black hole that could partly or totally impede the ejection of $^{56}$Ni and therefore the luminosity of the associated SN emission. While this sequence of event may be somewhat speculative it stresses the point that all the connections we detected and related classifications are related to the central engine, its formation and evolution.

Flares, that have been fully recognized and identified thanks to the Swift observations, are relevant in all of this since they give information about the activity of the central engine. Since the early observations [5, 6] we noticed, superimposed to an astonishingly standard X-ray light curve, the presence of flares that would occur in any type of burst, short and long, and at any cosmological age. While the occurrence of a burst may depend on various parameters related to the stellar evolution and may reach the final black hole phase either via the merging of two degenerated stars or following the collapse of a massive star, the flare activity is independent of any previous history and directly related to the burst and/or to the central engine. Flares are therefore not determined by the environment nor by the progenitor history.

Thanks to the many theoretical papers and analysis published on the subject [7, 8, 9] we seems to understand the basic features. An interesting detailed analysis has been given by Wu et al. [10] who concluded, with the data then available, that we may have flares due to internal shocks and others due to external shocks. Theoretical implication and observational evidence [11] [12], on the other hand, seem to favor an internal shock origin caused by an active engine. A slightly different point of view is discussed by Guetta et al. [13] where the main mechanism is due to new energy injection and "refreshing".

By the time of writing and after completion of the analysis of all the flares observed by XRT till January 31$^{st}$ 2006 (sample I) we completed part of the work so that for more general details that were presented at this meeting we refer to Chincarini et al.[14] and Falcone et al. [15].

## 2. Flare decay slope and reference time

For a source of radiation emitting isotropically and moving at speed $\beta = \frac{v}{c}$ with respect to the observer's frame, the Doppler factor is $D = \frac{1}{\gamma(1-\beta\cos\theta)}$ [16] ($\theta$ is the relativistic beaming opening angle). The Doppler factor for large Lorentz factor: $\gamma \gg 1$, $\beta \sim 1 - \frac{1}{2\gamma^2}$, $\cos\theta = 1 - \frac{\theta^2}{2}$ and therefore $D = \frac{1}{\gamma\left(1-\frac{v}{c}\cos\theta\right)} \approx (for\ v > 0)\ 2\gamma$.

Assuming a power spectrum of the emission (co-moving $I(\nu') \propto [\nu']^{-\beta}$ and $\nu = D\nu', d\Omega = D^{-2} d\Omega'$) and using the relativistic invariant, we have for the flux in the band $\Delta\nu$:

$$F(\Delta\nu) = \frac{Energy}{cm^2\ s} = I(\nu)\Delta\nu\ \Delta\Omega = I(\nu')\frac{\nu^3}{\nu'^3}\Delta\nu\ \Delta\Omega =$$

$$I(\nu')D^3\ \Delta\nu'D\ \Delta\Omega'D^{-2} = (\nu')^{-\beta}D^2\Delta\nu'\ \Delta\Omega' = D^2 F'(\Delta\nu')$$

Since $dt = dt' D^{-1}$ we also have $F(\Delta v) = (v)^{-\beta} D^{\beta+2} \propto (v)^{-\beta} t^{-2-\beta}$ and the well known result for the curvature, $\alpha = 2+\beta$ [17] [9] where $\alpha$ is defined by the equation of the observed flux: $F_v \propto v^{-\beta} t^{-\alpha}$. The maximum decay slope we can observe, that is the slope the observer measure if the source is switched off suddenly, is $\alpha = 2+\beta$.

The time must be measured from the onset of the event. The $T_0$ problem in the estimate of the decay slope was discussed at length especially in Chincarini et al. [18]. The same reasoning used for the early decay slope of the "standard" afterglow light curves applies to the flares light curves as well. Liang et al.[7] in particular use the maximum decay slope ($\beta \sim 1$) to estimate $T_0$ and find that the required $T_0$ is, for most cases, at the beginning of the flare.

To test the decay slope, that as we will see later is always well fit by a power law, we selected the time $T_0$ as the time at which the observed flux of the flare was 1% of the peak flux. In a smaller sample, in which both the rising and falling slopes could be fit and measured separately and reasonably well, we estimated the slope (to evaluate better the role of $T_0$) also with $T_0$ defined as the time at which the flux of the flare was about 5% of the peak intensity. The light curves of the flares have been always extracted from the observed light curve by fitting the underlying standard light curve with a multi break, generally 2 or 3, power law. Such operational procedure implies a model however. Observations evidence indeed that flares are superimposed to a standard underlying curve that maintains the characteristics observed in the afterglow of GRBs in which any large flare activity is absent. Flares, furthermore, do not perturb after their occurrence the shape of the underlying light curve showing that the two phenomena are unrelated to each other.

### 3. Flare decay slope and morphology

The early giant flare observed in GRB050502B [5, 19] remains a good reference for the morphology. After subtracting a power law underlying afterglow the profile of the flare is as shown in Figure 1. Such a shape is characteristics of many flares while others, as shown in Chincarini et al.[14], have a power law rising profile. This morphology is similar to the profile derived from simulations [20, 21] and coincides with the profiles observed in prompt emission pulses. The flare can be fitted, however, not only by the expression $a(1 - b x^{-c})$ + simple power law or power law + power law, but also using the relation given by Kobayashi et al.[22], that is:

$$F(t) = \begin{cases} 0 & t < 0 \\ h\left(1 - \dfrac{1}{(1+t)^2}\right) & t < 1 \\ h\left\{\dfrac{1}{[t^m]} - \dfrac{1}{(1+t)^2}\right\} & t > 1 \end{cases}$$

where $h = \text{Peak flux}/0.75$, the time is in units of $\frac{\delta t_e}{2\gamma_m^2}$, the time at which the reverse shock crosses the rapid shell, and $\gamma_m$ the Lorentz factor of the merged shells.

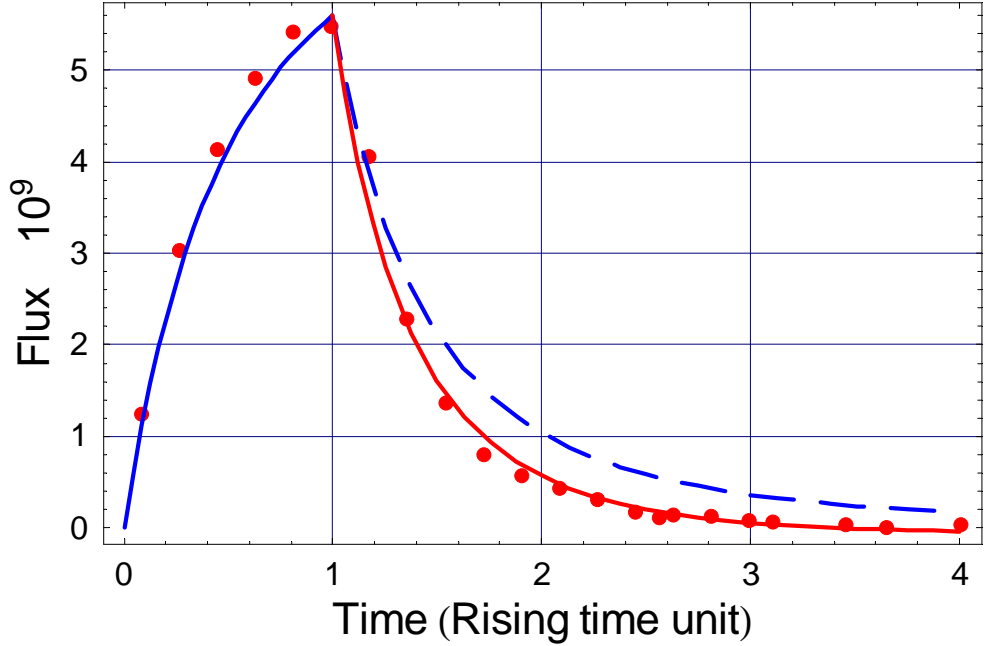

*Figure1. The peak flux observed in this flare that occurred in GRB050502B and start at $T_0 = 538.6$ s (defined as the time at which the emission was 5% of the peak flux) is of 5.6 $10^{-9}$ erg cm$^{-2}$s$^{-1}$. The dashed curve has been plotted using the equation by Kobayashi et al. without any adjustment (m=2) while the continuous curve is the best fit (m=2.42 ± 0.32).*

In the equation given by Kobayashi et al. m =2, dashed curve in Figure 1. The decline is much faster however and the best fit, leaving m as a free parameter in the above equation, gives m = 2.42 ± 0.32. The simple morphology observed and the good fit we obtained using the theory developed for the internal shocks argue in favor of a mechanism similar to that producing the prompt emission flares. The GRB050502B was here illustrated not only because this was the first very good example we analyzed, but also because it is one of the flares that show the steepest decay we measured. The slope we measure on this normalized light curve with a $T_0$ defined as above is α = 3.21 ± 0.77. This is perfectly in agreement with the curvature naked burst paradigm especially if we account for a rather soft spectral index: β ~ 2. This single burst gives us the following basic information valid both for the standard underlying light curves and for the flares: a) the flare is due to internal shock and b) the fast decay problem was a false problem and simply due to the wrong use of $T_0$.

Finally the functional form of the equation used for the fit is, to some extent, a choice that is little related to the physical model. The decay curve, for instance, is fitted quite well by a power law (used below for the statistical sample) and also using the exponential form used by Norris et al. [23]. In this case the falling flare profile (not normalized) is fitted by the curve $F(t) = F_{Peak} e^{-\left(\frac{t-t_{Peak}}{\sigma}\right)^m}$ with σ = 92±22 s (a measure of the width) and m = 1.12 (a measure of the sharpness of the pulse).

These indications have been fully confirmed by the statistical analysis of all the flares observed by XRT from launch to January 31st. Early in the analysis of this sample it was decided to choose the $T_0$ as the time at which the flare flux is 1% of the peak intensity. The rational was the best compromise to an operational definition to have the maximum counts possible for the study of the spectroscopic evolution [15]. This reference time is somewhat ill because of noise and the time of the peak is generally more accurate so that eventually that could be used as a reference for future work comparing models and observations. On the other hand a few flare, with good signal to noise ratio, have been measured in various ways so that it is fairly easy to understand the effect of the definition.

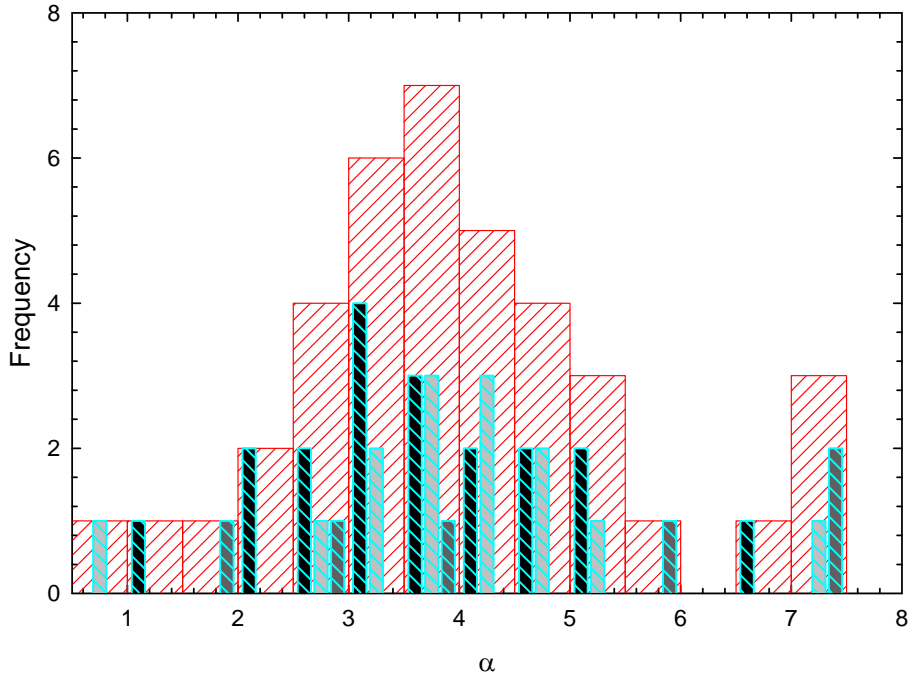

*Figure 2. Frequency of the temporal flare light cure decay index α. The main histogram dashed from right to left gives the distribution of all the measured flares while the different histograms dashed from left to right (descending) are the distribution of the first flare for each burst, for the second when present and for the third one. The peak of the distribution of the first flare seems to occur at smaller α than that for the distribution of the second flare.*

Always using GRB050502B main flare, as an example, we compute a slope of $\alpha = 6.3$ defining $T_0$ as the time at which the flux of the flare is 1% of the Gaussian peak while we obviously measure a smaller value selecting the time at which the flux is 5% of the Gaussian peak and $\alpha = 3.21 \pm 0.77$ ($\alpha = 1.17$ assuming $T_0 = T_{Peak}$) if we normalize the light curve as described above. Accounting for the fact that the absolute value is model dependent and that in some case we may have the presence of an undetectable blends of flares that tends to decrease the slope, the measured distribution is shown in Figure 2.The histograms with narrow bars, decreasing dashing lines from left to right, refer to the first, second and third flare observed on a given GRB. While we have almost no statistics for the distribution of the third flare, the distribution of the firsts and seconds are very similar. The main histogram with large bars, decreasing dashing lines from right to left, gives the distribution of all the observed flares, it is the sum for each bin, of the previous distributions. For the three distributions, FIRST, SECONDS and TOTAL (this includes the third bursts that have not enough statistics to be considered separately) we have for $\alpha$:

|         | Mean | Median | Variance |
| --- | --- | --- | --- |
| FIRSTS  | 3.17 | 3.20 | 1.77 |
| SECONDS | 3.55 | 3.50 | 2.10 |
| TOTAL   | 3.45 | 3.35 | 2.35 |

### 4. The "afterglow" curve of the flares in GRB051117A and GRB060111A.

We begin to have a fairly good understanding on the underlying standard light curves. The shallow decay following the early steep decline phase is not yet fully understood since the energy injection mechanism suggested in previous work needs too much fine tuning to be realistic. Of particular interest, on the other hand, is the analysis carried out by Willingale et al. [24] where the shallow slope is simply due to the peak or plateau of the afterglow component and this is also in agreement with the early sketch put forward by Sari [25]. One of the striking features of the shallow phase is that it always occurs in a rather constant time interval after the prompt emission. This could naturally fit under the assumption that the circum-stellar medium of a massive star before collapse is more or less the same for all progenitors. On the other hand such shallow decay is observed also in the afterglow light curve of short bursts as in GRB051221A, Figure 3 top left. The circum-stellar environment of a short burst certainly differs from that of a long GRB. GRB050801, top right of Figure 3, is a possible example (there are many others) of a rising light curve after the early steep decay. This could be eventually also be explained by the presence of a small flare and however there is no clear cut between the two interpretations. To further evidence that often we may have an ambiguous interpretation on the bottom left of Figure 3 we show the light curve of GRB050814 where again rather than evidencing a small flare we prefer to use a triple broken power law. Is it only a matter of semantic? While also in these cases the model of a decaying prompt emission followed by an emerging afterglow seems reasonable, albeit what we mentioned above, we may have a further complication for the model due to subsequent bumps generally preceding a steeper slope.

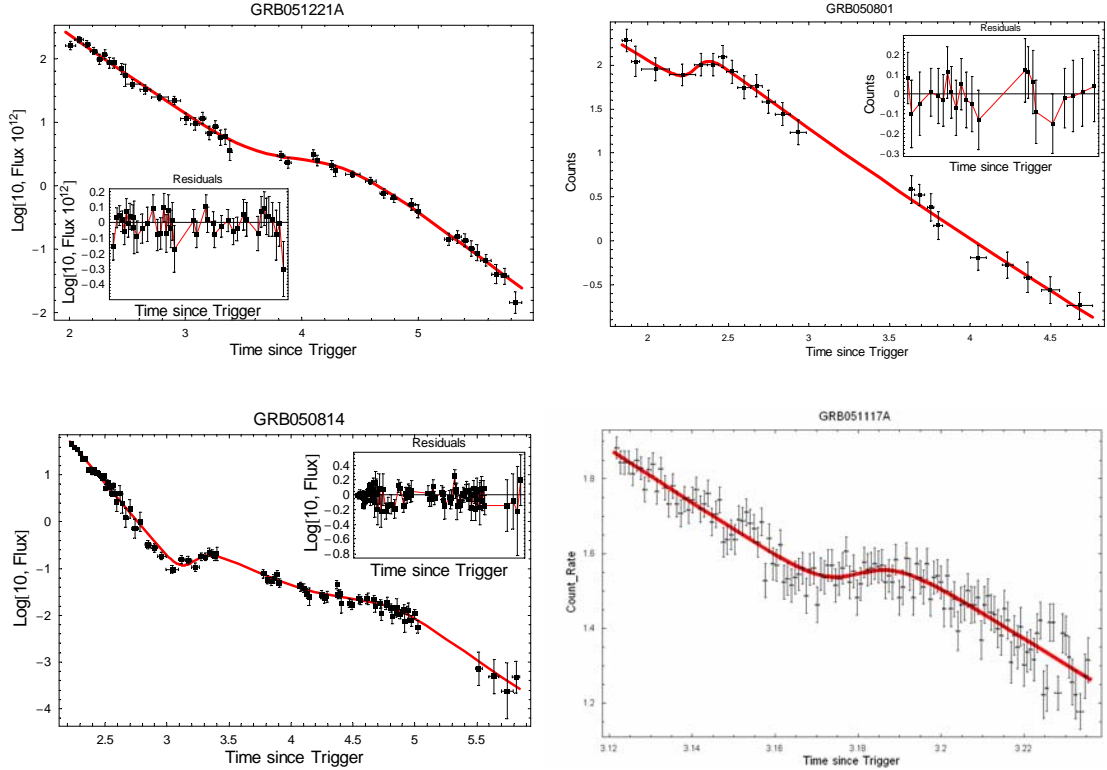

*Figure 3. Top left: XRT light curve of the short burst GRB051221A to evidence the broken power law; Top right: GRB050801 with a bump that could be equally well fitted by a broken power law (always due to the injection of some kind of energy) or with a small flare; Bottom left: an interesting example of a light curve fitted by a triple broken power law, again a possible alternative to the presence of a small flare; Bottom right: the decaying light curve of the flare in GRB051117A showing the same characteristics of a standard underlying light curve.*

The tail of the GRB051117A flare, Figure 3 bottom right, shows a striking similarity, pseudo-fractal behavior, with the standard afterglow light curve. Here the fit, always after subtraction of the underlying light curve, is also done with a broken power law. A similarity in the decaying light curve does not necessarily implies that the same physics is at work, it is a possibility however. For the flares it is reasonably easy to envisage bumps and blends due either to the collision of relativistic shells or to the variable activity of the central engine.

## 5. Width, time of occurrence and $\Delta t_{fall}/\Delta t_{rise}$

In addition to the morphology, parameters that depend directly from the mechanism generating the shock are the ratio $\frac{\Delta t}{t} = \frac{Width\ of\ the\ flare}{time\ of\ occurrence}$ and $\frac{\Delta t_{fall}}{\Delta t_{rise}} = \frac{time\ to\ rise\ to\ peak}{time\ to\ fall}$. Theoretical constraints on the ratio $\frac{\Delta t}{t}$ have been given by

Wu et al. [10], Ioka et al. [27] and Zhang et al.[8] discussing various mechanisms capable of producing flares. Lazzati and Perna [12] more recently put tighter constraints and came also to the conclusion that the flares must be due to internal shocks. To measure this ratio we fitted underlying light curve and flares simultaneously using a Gaussian curve for the flares. For this purpose the Gaussian fit is a good fit and at the same time allows an unbiased measure of the width maximizing the statistics (a width and a maximum can be measured also on flares defined by few data points). The distribution of the ratio, as shown at the Venice meeting, is given in Figure 4.

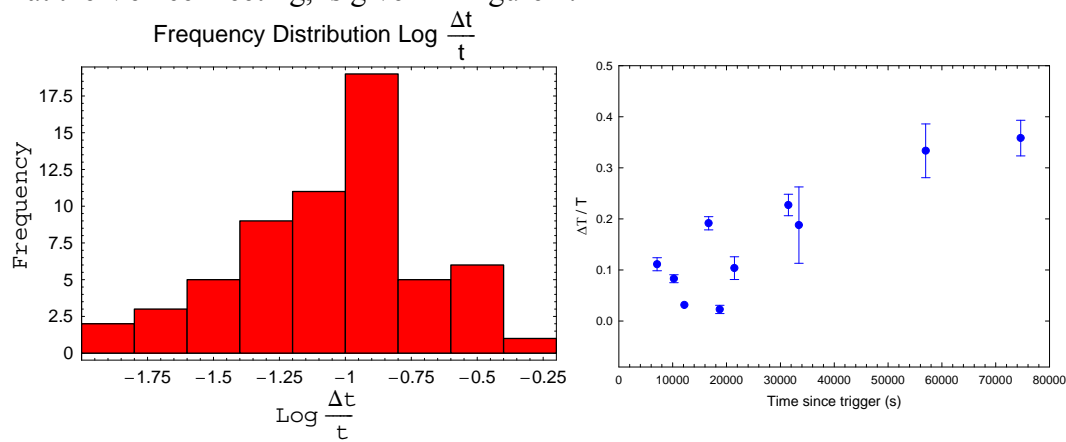

Figure 4. Left: Distribution of the width to peak time ratio for the whole sample; Right: $\frac{\Delta t}{t}$ for the flares observed long after the BAT trigger time. The two last ratios have been measured in GRB050502B (long GRB) and GRB050724 (short GRB).

The mean value $<\frac{\Delta t}{t}>$ = 0.12 with a standard deviation = 0.10 tend to exclude an external shock origin. At the time of the meeting we considered as the most plausible explanation the dissipation due to collisions of shells released by the central engine during the prompt emission phase ("lazy shells"). Lazzati and Perna, on the other hand, point out that in this case we would expect $\frac{\Delta t}{t} \geq 0.25$ while the distribution of Figure 4 left is skewed toward smaller values with a mean value that is also smaller. It seems therefore that the most likely origin is due to shells that are ejected after a time that is much larger than the timescale of the prompt emission. On the right of Figure 4 we plotted the group of late flares that occurred more than 10000 s after the trigger time. The tendency to a larger value of $\frac{\Delta t}{t}$ is dominated by the two last measurements. Furthermore a few early flares have a similar large $\frac{\Delta t}{t}$ value so that we can not state that this is a characteristic of late flares. This is naturally a key issue and it will be revisited shortly also using the larger sample that we now have.

Norris et al. [23] in their study of the BATSE pulses noticed that the most frequently occurring decay-to-rise ratio is about 2.5. In our sample we measure $\left\langle \frac{\tau_D}{\tau_R} \right\rangle = 2.35$ with a standard deviation $\sigma = 1.71$. There is a tendency, Figure 5, in the data for the early flares to show a correlation of $\frac{\tau_D}{\tau_R}$ with $T_{90}$ (flare) (the width of the flare measured at the 5% of the peak flux). However the only two late flares we could measure: that in GRB050730 occurring at t = 4589 s and the late flare of GRB050724 occurring at t = 54954 s are largely off the correlation. Further data will clarify the behavior of the late flare, nonetheless we have again an indication that some of the attributes of the flares are very similar to the attributes measured in the prompt emission pulses.

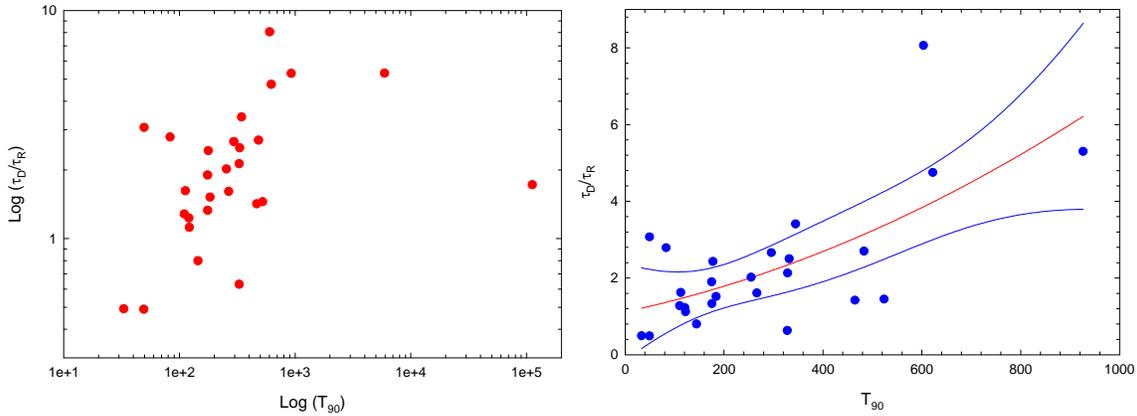

Figure 5. $\frac{\tau_D}{\tau_R}$ as a function of $T_{90}$. $\tau_D$ is the time elapsed during the decay between the peak of the flare and the 5% of the peak flux, $\tau_R$ the time elapsed during the rising of the curve between the time the flux if 5% of the peak and the flux of the peak. The rising and falling light curves have been fitted separately using either the expression $a(1 - b\, x^{-c})$ or a simple power law. The peak in this case has been measured as the time at which the two fits crosses and $T_{90}$ (flare) = $\tau_D + \tau_R$. The continuous lines are the best fit and the 95% confidence level.

## 6. Conclusions

This first sample of GRBs tend to favor the origin of flares as due to the collision of ultra-relativistic shells ejected by the central engine after, and in some cases long after, the prompt emission phase. The characteristics of a few flares would also agree with the collision of low Lorentz factor shells emitted during the prompt emission phase while it is almost completely excluded, based on current theoretical formulations, the possibility that such flares are due to external shock. As previously noticed following the observations of flares also in the short GRB050724, the origin of flares must be unrelated to the characteristics of the circum-stellar medium and progenitor and depends solely on the final common product, i.e. the accretion disk. Clearly this fairly commonly accepted scenario is strongly model dependent and somewhat related also to the possibility of defining a third class of GRBs (prototype GRB060614).

These findings have strong implications for the activity of the central engine. The suggestion is that following a prompt emission activity likely due to the ejection of shells with unknown Lorentz factor distribution the following activity is sparse and generally limited to a smaller number of shells. At the same time the energetic is critical. The largest flares observed have a fluence that is of the same order of magnitude or larger (GRB050502B) than the fluence of the underlying afterglow. Accounting for the fact that the efficiency of the internal shock is of about 1% [27] the energy involved in a single flare is extremely high. Late flares, in addition, are equally energetic and however differ from early flares since they last longer and have smaller peak intensity. This would suggest thicker shells and likely favor the early ejection. This matter will be solved as soon as we have more statistics and complete modeling. Finally we have the mini-variability, i.e. the presence of flares of very small amplitude. The characteristic and the origin of these fluctuations is still under investigation.

REFERENCES

[1]     Gehrels, N. et al., 2004, Ap.J. 611, 1005
[2]     Gehrels, N., 2006, These proceedings
[3]     Norris, J.P. and Bonnell, J.T. 2006, Ap.J., 643, 266
[4]     Della Valle et al. 2006, Nature, 444,1050
[5]     Burrows, D. N. et al., 2005, Science, 309, 1833
[6]     Romano et al., 2006, A&A 450, 59
[7]     Liang, et al., 2006, Ap.J. submitted, astro-ph/0602142
[8]     Zhang B. et al., 2006, Ap. J. 642, 354
[9]     Dermer, C. D., 2004, Ap. J. 614, 284
[10]    Wu, X. F. et al., 2005, Ap. J. submitted, astro-ph/0512555
[11]    Chincarini,G., 2006, astro-ph/0608414 & Chincarini,G. et al., 2006, astro-ph/0612121
[12]    Lazzati D. & Perna R. 2006, MNRAS, submitted, astro-ph 0610730
[13]    Guetta, D. et al. These proceedings
[14]    Chincarini, et al., 2007, in preparation
[15]    Falcone, A.D. et al., 2007, in preparation
[16]    Rybicki, G.B, & Lightman, A.P. 1979, Radiative Processes in Astrophysics
[17]    Kumar, P. & Panaitescu, A., 2000, Ap. J., 541, L51
[18]    Chincarini, G. et al., 2005, astro-ph/0506453
[19]    Falcone, A. D., 2006, Ap.J. 641,1010
[20]    Daigne, F. & Mochkovitch, 1998, MNRAS 296, 275
[21]    Shen, R-F et al., 2005, MNRAS 362, 59
[22]    Kobayashi, S., Piran, T. & Sari, R., 1997, Ap. J. 490, 92
[23]    Norris, J.P. et al., 1996, Ap. J. 459, 393
[24]    Willingale R. et al., 2006, submitted to Ap. J., astro-ph/0612031
[25]    Sari, R., 1997, Ap. J. 489, L37
[26]    Ioka, K. et al., 2005, Ap. J. 631, 429

[27] Kumar, P., 1999, Ap. J. 523, L113